# Interlayer exchange coupling through Ir-doped Cu spin Hall material

Hiroto Masuda[1,*], Takeshi Seki[1,2,†], Yong-Chang Lau[1,2], Takahide Kubota[1,2], and Koki Takanashi[1,2,3]

[1] Institute for Materials Research, Tohoku University, Sendai 980-8577, Japan

[2] Center for Spintronics Research Network, Tohoku University, Sendai 980-8577, Japan

[3] Center for Science and Innovation in Spintronics, Tohoku University, Sendai 980-8577, Japan

E-mail: * masuda1610@imr.tohoku.ac.jp, † go-sai@imr.tohoku.ac.jp

## Abstract

Metallic superlattices where the magnetization vectors in the adjacent ferromagnetic layers are antiferromagnetically coupled by the interlayer exchange coupling through nonmagnetic spacer layers are systems available for the systematic study on antiferromagnetic (AF) spintronics. As a candidate of nonmagnetic spacer layer material exhibiting remarkable spin Hall effect, which is essential to achieve spin-orbit torque switching, we selected the Ir-doped Cu in this study. The AF-coupling for the Co / $Cu_{95}Ir_{5}$ / Co was investigated, and was compared with those for the Co / Cu / Co and Co / Ir / Co. The maximum magnitude of AF-coupling strength was obtained to be 0.39 mJ/m$^2$ at the $Cu_{95}Ir_{5}$ thickness of about 0.75 nm. Furthermore, we found a large spin Hall angle of $Cu_{95}Ir_{5}$ in Co / $Cu_{95}Ir_{5}$ bilayers by carrying out spin Hall magnetoresistance and harmonic Hall voltage measurements, which are estimated to be 3 ~ 4 %. Our experimental results clearly indicate that $Cu_{95}Ir_{5}$ is a nonmagnetic spacer layer allowing us to achieve moderately strong AF-coupling and to generate appreciable spin-orbit torque via the spin Hall effect.

# 1. Introduction

In the research field of spintronics, antiferromagnetic (AF) materials have recently been taken much notice of [1–3] thanks to low magnetic susceptibility [4], lack of magnetic stray field, and fast magnetization dynamics [5–8]. Although those characteristics fascinate us largely, one of the major challenges of AF spintronics is to establish a method to manipulate the magnetic moments of the AF structure efficiently because of the difficulty in the control of their magnetic moments by applying an external magnetic field due to low magnetic susceptibility. Recently, several works reported the interaction between AF structures and electric current [3,9–17]. The electrical switching in CuMnAs, which is one of the representative materials for AF spintronics, was demonstrated using current-induced internal fields originating from its crystal structure with broken inversion symmetry [9]. Nickel oxide (NiO) is also a bulk AF material, and it was reported that the alignment of antiferromagnetically coupled magnetic moments, i.e. Neel vector, could be switched by the spin-orbit torque (SOT) resulting from the spin Hall effect (SHE) in an adjacent Pt [10]. Particularly, the SOT is a promising technique for the switching of AF structure since SOT is generated simply by passing an electric current through a spin Hall material *e.g.* Pt, Ta, and W [18–24].

Most of the research efforts in AF spintronics have focused on bulk AF materials. For the bulk AF materials, however, the complicated AF domain structures are often observed [10,16,17], and AF-coupling strength are the intrinsic properties and are almost uncontrollable. These features are obstacles for the systematic study on SOT in the AF structure. Another candidate system for studying AF spintronics is a metallic superlattice showing an AF structure, in which the ferromagnetic layers separated by the nonmagnetic spacer layer are antiferromagnetically coupled through interlayer exchange coupling [25,26]. In contrast to the bulk AF materials, antiferromagnetically-coupled metallic superlattices allow us to control their AF-coupling strengths by changing the layer thicknesses [27–39]. Moreover, many interfaces in the metallic superlattice have a possibility to lead to the enhancement of SOT. For example, the enhanced field-like SOT at the Pt interface was reported theoretically [40] and experimentally [41]. In addition, the first-principles calculation predicted a giant interface spin Hall angle ($\theta_{SH}$) between Pt and $Ni_{80}Fe_{20}$ [42]. Considering the advantages mentioned above, antiferromagnetically-coupled metallic superlattices are promising systems for the systematic investigation of the SOT on the AF structure if one can find a nonmagnetic spacer layer material simultaneously exhibiting sufficient interlayer exchange coupling and large SHE.

This study focuses on the metallic superlattice with an Ir-doped Cu (Cu-Ir) nonmagnetic spacer layer sandwiched by Co ferromagnetic layers. The Co / Cu / Co and Co / Ir / Co systems are well-known materials combinations which show the interlayer exchange coupling and giant magnetoresistance (GMR) [27–34,37–39,43]. Thus, the Cu-Ir nonmagnetic spacer layer is expected to show AF-coupling, and may allow us to electrically detect the magnetization orientation through the GMR effect. In addition, Cu-Ir is a promising spin Hall material. Although the nonmagnetic Cu and Ir show negligibly small SHE [44], Ir-doped Cu was reported to exhibit finite extrinsic SHE [45–48], and the resultant $\theta_{SH}$ of the Cu-Ir was 2.1 ± 0.6 (%) [45]. Therefore, one may anticipate that the Co / Cu-Ir / Co superlattice is a promising system to simultaneously show the AF-coupling and the SHE.

In this paper, we report the systematic investigation of the AF-coupling strength for epitaxially grown Co / $Cu_{95}Ir_5$ / Co tri-layers, and compare the magnetic properties between Co / $Cu_{95}Ir_5$ / Co, Co / Cu / Co, and Co / Ir / Co systems. In addition to the investigation of AF-coupling behavior, we evaluate $\theta_{SH}$ of the present $Cu_{95}Ir_5$ using two different experimental methods: spin Hall magnetoresistance (SMR) measurement and harmonic Hall voltage measurements. We also discuss the relationship between interlayer exchange coupling and SHE qualitatively.

## 2. Experimental procedure

All the thin films were deposited on $Al_2O_3$ (0001) single crystal substrates using an ultrahigh vacuum magnetron sputtering system with a base pressure of the order of $10^{-7}$ (Pa). The stacking structure of the samples for structural characterization and magnetization measurements is as follows: $Al_2O_3$ (0001) subs. / Cr (10 nm) / Au (5 nm) / Cu (35 nm) / Co (2 nm) / X ($t$) / Co (2 nm) / Cu (2 nm) / Cr (5 nm), where X = Cu, $Cu_{95}Ir_5$ or Ir. In this study, the samples with X = Cu, $Cu_{95}Ir_5$ and Ir are named Cu interlayer sample, $Cu_{95}Ir_5$ interlayer sample and Ir interlayer sample, respectively. The 10 nm-thick Cr underlayer was annealed at 600 °C in order to obtain a flat surface. The annealing at 800 °C was also performed after depositing the 35 nm-thick Cu underlayer, which led to the alloying of Au and Cu underlayers. The remaining stacks were deposited at ambient temperature. The $Cu_{95}Ir_5$ layer was prepared by co-sputtering Cu and Ir elemental targets. The composition of $Cu_{95}Ir_5$ was determined by inductively coupled plasma (ICP) analysis. Magnetic properties were measured using a vibrating sample magnetometer (VSM) for Cu and $Cu_{95}Ir_5$ interlayer samples, and a superconducting quantum interference device (SQUID) magnetometer for Ir interlayer sample. Atomic force microscopy (AFM) enabled us to characterize the surface morphology and the surface roughness. Structural characterization was carried out using reflection high energy electron diffraction (RHEED) and out-of-plane x-ray diffraction (XRD) with Cu-K$\alpha$ radiation.

The transport properties were characterized using four-terminal configurations in a physical property measurement system (PPMS, Quantum Design Japan, Inc.). The films were patterned into a Hall bar structure with a nominal length ($l$) of 50 μm and a width ($w$) of 10 μm through the use of photolithography and Ar ion milling. As the electrodes and the contact pads, Cr (20 nm) / Au (200 nm) was deposited using ion beam sputtering. The GMR curves were measured for the Cu and $Cu_{95}Ir_5$ interlayer samples to confirm the AF-coupling electrically as well as the evaluation of GMR ratio. Co / $Cu_{95}Ir_5$ bilayers without the seed layers were also prepared in order to evaluate $\theta_{SH}$ and the spin diffusion length ($\lambda_{SD}$) of $Cu_{95}Ir_5$. The stacking structure of the thin film for evaluating SHE is as follows: $Al_2O_3$ (0001) subs. / Co (2 nm) / $Cu_{95}Ir_5$ (*t* nm) / Cr (5 nm). Both the SMR and harmonic Hall voltage measurements were employed for determining $\theta_{SH}$ and $\lambda_{SD}$ of $Cu_{95}Ir_5$. For GMR and SMR measurements, a direct current (dc) of 100 μA was applied. For harmonic Hall voltage measurements, an alternate current (ac) of amplitude 5 mA and frequency 13.7 Hz was applied using a Keithley 6221 dc and ac current source meter (Tektronix, Inc.). The first and second harmonic Hall resistances were simultaneously measured by two LI5660 lock-in amplifiers (NF Corporation). All the measurements were done at 300 K with an external magnetic field (*B*) applied in the PPMS.

## 3. Results and discussion

### 3.1. Structural characterization

Figure 1(a) shows the AFM image for the buffer layer consisting of Cr (10 nm) / Au (5 nm) / Cu (35 nm). The average surface roughness ($R_a$) was obtained to be $R_a \sim 0.10$ nm, which suggests the formation of a flat buffer layer. Figures 1(b), 1(c), and 1(d) display the RHEED patterns taken after the growth of top Co layers for the Cu, $Cu_{95}Ir_5$, and Ir interlayer samples, respectively. Clear streak patterns were observed and the patterns changed with the substrate rotation for all the samples, implying that all the Co / X / Co layers were epitaxially grown on the $Al_2O_3$ (0001) single crystal substrate.

The structures of the buffer layer, Cu, $Cu_{95}Ir_5$, and Ir interlayer samples were characterized by out-of-plane XRD. The results are shown in Fig. 1(e), in which the expected peak angles based on the powder diffraction data are also as shown in the top figure. In the XRD profile for the buffer layer, only the peaks at $2\theta \sim 42°$ and 44.5° are observed, which are identified as Cu-Au 111 and Cr 110, respectively. Neither reflection from Cu nor Au and the appearance of Cu-Au peak indicate that the Au / Cu layers were completely alloyed by annealing at 800°C. For the samples with the Co layers, namely the Cu, $Cu_{95}Ir_5$, and Ir interlayer samples, an increase in the background could be found around the Co 111 peak position compared to the XRD profile for the buffer layer, which is attributable to the appearance of Co 111 peak. These XRD profiles suggest that the (111)-oriented Co layer was grown on the Cu-Au buffer layer.

*3.2. Magnetic properties*

Figures 2(a), 2(b), and 2(c) show the magnetization curves for the Cu ($t$ = 0.75 nm), $Cu_{95}Ir_5$ ($t$ = 0.75 nm) and Ir ($t$ = 0.5 nm) interlayer samples, respectively. $B$ was applied in the film plane for all the samples. The small remanent magnetization ($M_r$) and large saturation field ($B_s$) are observed for the $Cu_{95}Ir_5$ interlayer sample as well as the Cu and Ir interlayer samples, indicating that the magnetizations of two Co layers are aligned antiferromagnetically even for the $Cu_{95}Ir_5$ interlayer sample. The values of $B_s$ were estimated to be 0.57 T for the Cu interlayer sample, 0.33 T for the Cu-Ir interlayer sample, and about 7 T for the Ir interlayer sample. A possible reason for non-zero $M_r$ is the coexistence of both AF-coupling region and non-AF-coupling region such as ferromagnetic coupling or 90° coupling due to the magnetic-dipole field or the fluctuation of thickness originating from the layers roughness [49–52]. We highlight that AF-coupling-like behavior is observed even for the Ir-doped Cu interlayer. Since the Ir interlayer sample gives rise to much larger $B_s$ and much smaller $M_r$ than those for the Cu interlayer sample, one might expect the enhanced AF-coupling when Ir is doped into Cu. Actually, however, $B_s$ ($M_r$) was decreased (increased) when 5 at.% Ir was doped into Cu.

The AF-coupling strength ($J_{AF}$) was estimated from the magnetization curves. $J_{AF}$ is given by $J_{AF} = -(1/2) t_{Co}M_sB_s$, where $t_{Co}$ is the thickness of one Co layer and $M_s$ is the saturation magnetization, with the assumption that $B_s$ is much larger than the anisotropy field within the plane [53]. Figures 3(a), 3(b), and 3(c) plot the $t$ dependence of $M_s$, $B_s$, and - $J_{AF}$, respectively, for the Cu (red circles), $Cu_{95}Ir_5$ (blue squares), and Ir (purple triangles) interlayer samples. For all the samples, $M_s$ keeps almost the constant value of 1300 kA/m regardless of $t$. On the other hand, $B_s$ and $-J_{AF}$ are clearly varied depending on $t$. These results are qualitatively interpreted as the characteristic feature for the interlayer exchange coupling. The maximum $-J_{AF}$ values ($-J_{max}$) were obtained to be 0.76 mJ/m$^2$ at $t$ = 0.75 nm for the Cu interlayer sample, 0.39 mJ/m$^2$ at $t$ = 0.75 nm for the $Cu_{95}Ir_5$ interlayer sample, and 8.12 mJ/m$^2$ at $t$ = 0.5 nm for the Ir interlayer sample. It has been found that the AF structure is formed by interlayer exchange coupling through the $Cu_{95}Ir_5$ interlayer with 0.6 nm < $t$ < 1.0 nm. The values of $-J_{max}$ reported previously were 0.3 ~ 1.3 mJ/m$^2$ for the Co / Cu / Co systems with $t$ = 0.7 ~ 0.9 nm [27–31,37–39], and 2.04 mJ/m$^2$ at room temperature, 2.05 mJ/m$^2$ at 10 K, and 6 mJ/m$^2$ at 4.2 K for the Co / Ir / Co systems with $t$ = 0.5 ~ 0.6 nm [33,34]. One sees that the present Co / Cu / Co sample possesses the $-J_{max}$ value comparable to the previous values and the present Co / Ir / Co sample shows the $-J_{max}$ value 4 times larger than the previous ones measured at room temperature. This fact implies that $-J_{max}$ may be enhanced by the high crystallinity and/or the well-defined interfaces for the present samples thanks to the epitaxial growth on the single crystal substrate. It is noted that the values of $-J_{AF}$ for the $Cu_{95}Ir_5$ interlayer sample was smaller than the Cu and Ir interlayer samples. This means that AF-coupling does not monotonically increase when Ir is doped into the Cu interlayer even though the Ir interlayer sample exhibits the large $-J_{AF}$. According to the Ruderman-Kittel-Kasuya-Yoshida (RKKY) exchange model extended by Bruno and Chappert [54], the spacer layer thickness showing the indirect exchange coupling is determined by the wave vectors ($\mathbf{q}_s$) linking two points of Fermi surface with antiparallel velocities in the case of noble metal spacers such as Cu, Ag, and Au. Although this study has not observed the oscillatory behavior of interlayer exchange coupling, which is theoretically considered to relate with $\mathbf{q}_s$ [54], the fact that a peak in $-J_{max}$ was obtained at $t$ ~ 0.75 nm for both the Cu and $Cu_{95}Ir_5$ interlayer samples implies the possibility that doping 5 at.% Ir into Cu does not remarkably change the electronic structure of the pure Cu.

*3.3. Giant magnetoresistance (GMR)*

Measuring the GMR effect is another way to confirm the AF-coupling for the Co / X / Co samples. Figure 4(a) shows a schematic illustration of the device and the coordinate system used for the MR measurement including GMR and SMR. The longitudinal resistance ($R_{xx}$) was detected by applying an electric current ($I_c$) of 100 μA and $B$ along the $x$ direction ($B_x$). Figure 4(b) shows the optical microscope image of a Hall bar together with the terminal configuration for the GMR measurement. The resultant MR curves for the Cu and $Cu_{95}Ir_5$ interlayer samples with $t$ = 0.75 nm are shown in Figs. 4(c) and 4(d), respectively. For both samples, the value of $R_{xx}$ decreases with increasing $B_x$ and saturates at $B_x \approx B_s$, This $B_x$ dependence of $R_{xx}$ is interpreted as the antiparallel alignment of $M$ between two Co layers at $B_x$ = 0 T and the parallel alignment at high $B_x$, which is consistent with the magnetization reversal process observed in the magnetization curves. Consequently, the clear GMR effect is seen for the $Cu_{95}Ir_5$ interlayer sample as well as the Cu interlayer sample. The appearance of GMR effect is an evidence for the spin-dependent transport. The GMR ratio is defined as $\{R_{xx}^{GMR}/R_{xx}^0\} \times 100 = [\{R_{xx}(B_x = 0) - R_{xx}(B_x = 8 \text{ kOe})\}]/R_{xx}(B = 0)] \times 100$. The GMR ratios were obtained to be 1.68 % for the Cu interlayer sample, and 0.64 % for the $Cu_{95}Ir_5$ interlayer sample. A reason for the reduction in the GMR ratio by doping 5 at.%-Ir into Cu is the imperfect AF alignment of Co layer magnetizations due to the reduced -$J_{AF}$ for the same sample. Another explanation is the shortened spin diffusion length in the $Cu_{95}Ir_5$ as discussed later. The appearance of GMR also confirms that the low $M_r$ values observed in the magnetization curves do not result from the formation of multiple magnetic domains but originate from the AF alignment of two Co layers magnetizations.

*3.4. Spin Hall magnetoresistance (SMR)*

As mentioned in Secs. 3.2 and 3.3, we have confirmed that Co / $Cu_{95}Ir_5$ / Co tri-layer shows appreciable AF-coupling. Hereafter, we evaluate the spin Hall effect for the present $Cu_{95}Ir_5$ alloy using two different methods: SMR measurement and harmonic Hall voltage measurement. The SMR is caused by the collective action of the direct and inverse SHE of a nonmagnetic (NM) layer adjacent to an electrically insulating magnet [18] or a ferromagnetic metal (FM) [21]. SMR can be utilized in order to estimate $\theta_{SH}$ and $\lambda_{SD}$ of the NM layer [21]. In this study, Co / Cu$_{95}$Ir$_5$ bilayers were employed for the SMR measurement, which have the following layer stacking: Al$_2$O$_3$ (0001) substrate / Co (2 nm) / Cu$_{95}$Ir$_5$ ($t_{Cu\text{-}Ir}$) / Cr (5 nm).

Figure 5(a) shows the $R_{xx}$ as a function of $B$ along the $y$ axis ($B_y$, green curve) and $z$ axis ($B_z$, blue curve) for Co (2 nm) / Cu$_{95}$Ir$_5$ (2 nm). The SMR ratio was defined as $\{\Delta R_{xx}^{SMR}/R_{xx}^0\} \times 100 = [\{R_{xx}(B_y) - R_{xx}(B_z)\}/R_{xx}(B=0)] \times 100$. The background slopes in high $B_y$ and $B_z$ regions were corrected from the raw data in order to remove the contributions of the ordinary magnetoresistance and the forced effect. The $\Delta R_{xx}^{SMR}$ corresponds to the resistance difference as denoted by the red arrow in Fig. 5(a).

The inverse of the sheet resistance $1/R_{xx} \cdot (l/w)$ is plotted as a function of $t_{Cu\text{-}Ir}$ in Fig. 5(b), when $I_c$ was set to 100 μA with no $B$. $t_{Cu\text{-}Ir}$ was varied in the range from 1.5 nm to 6 nm. The resistivity of Cu$_{95}$Ir$_5$ ($\rho_{Cu\text{-}Ir}$) was estimated to be 92.42 μΩ cm from the slope of the linear fit to the experimental data.

Based on a drift-diffusion model, the value of SMR of $\Delta R_{xx}^{SMR}$ for the Co / Cu$_{95}$Ir$_5$ bilayer is given by the following equation [21]:

$$\Delta R_{xx}^{SMR} \approx -\theta_{SH}^2 \frac{\lambda_{SD}}{t_{Cu-Ir}} \frac{\tanh^2\left(\frac{t_{Cu-Ir}}{2\lambda_{SD}}\right)}{1+\xi} \times \left[\frac{g_R}{1+g_R \coth\left(\frac{t_{Cu-Ir}}{\lambda_{SD}}\right)} - \frac{g_F}{1+g_F \coth\left(\frac{t_{Cu-Ir}}{\lambda_{SD}}\right)}\right],$$

$$g_R \equiv 2\rho_{Cu-Ir}\lambda_{SD}\text{Re}[G_{MIX}],$$

$$g_F = \frac{(1-P^2)\rho_{Cu-Ir}\lambda_{SD}}{\rho_{Co}\lambda_{Co}\coth\left(\frac{t_{Co}}{\lambda_{Co}}\right)}, \quad (1)$$

where, Re[$G_{MIX}$] is the real part of the spin mixing conductance, showing the absorption of spin current at the $Cu_{95}Ir_5$ / Co interface [55–58]. The parameters of $t_{Co}$, $\rho_{Co}$, $\lambda_{Co}$, and $P$ are the thickness, resistivity, spin diffusion length, and the current spin polarization of Co layer, respectively. The current shunting effect into the Co layer is taken into account by $\xi \equiv (\rho_{Cu\text{-}Ir} t_{Co} / \rho_{Co} t_{Cu\text{-}Ir})$.

The $t_{Cu\text{-}Ir}$ dependence of SMR ratio is shown in Fig. 5(c). The experimental data of SMR ratio were well fitted by Eq. (1) using the following parameters: $\rho_{Cu\text{-}Ir}$ = 92.42 μΩ cm and $\rho_{Co}$ = 52.62 μΩ cm, which were experimentally measured for the present $Cu_{95}Ir_5$ and Co, $Re[G_{MIX}]$ = 1.0 × $10^{15}$ $\Omega^{-1}m^{-2}$, $P$ = 0.30, and $\lambda_{Co}$ = 5.0 nm, which were chosen as typical values for the metallic bilayer systems with Co [59–63]. As a result, the two free parameters of the fit, $\theta_{SH}$ and $\lambda_{SD}$ of $Cu_{95}Ir_5$ were obtained to be 4.3 % and 1.3 ± 0.1 nm, respectively.

*3.5. Harmonic Hall voltage measurements*

In addition to the SMR measurements, the harmonic Hall voltage measurements were conducted for the Co (2 nm) / $Cu_{95}Ir_5$ ($t_{Cu\text{-}Ir}$) bilayers. Figure 6(a) illustrates the experimental setup for harmonic Hall voltage measurements together with the definition of the coordinates. The first and second harmonic voltages ($V^\omega$ and $V^{2\omega}$) were detected by applying an ac current with an amplitude of 5 mA and a frequency of 13.7 Hz along the *x* direction. The applied *B* was rotated in the *x-y* plane making an azimuthal angle ($\phi$) with respect to the *x*-axis. Since the applied *B* was sufficiently large to saturate the magnetization vector of Co layer (***M***), ***M*** totally followed the *B* direction and the $\phi$ coincides with the in-plane azimuthal angle of ***M***. Also, the polar angle of the magnetization vector ($\theta$) is π/2 because the Co layer is magnetized in the film plane.

According to Ref. [19,20], the first and second harmonic Hall resistances ($R^\omega$ and $R^{2\omega}$) are given by

$$R^\omega = R_{AHE} \cos\theta + R_{PHE} \sin^2\theta \sin(2\phi), \tag{2}$$

$$R^{2\omega} = \left[\left(R_{\text{AHE}} \frac{B_{\text{DL}}}{B + B_{\text{ani}}} + I_c \alpha \nabla T\right) \cos\phi + 2R_{\text{PHE}}(2\cos^3\phi - \cos\phi)\frac{B_{\text{FL}} + B_{\text{Oe}}}{B}\right], \quad (3)$$

where $R_{\text{AHE}}$ and $R_{\text{PHE}}$ are the resistance changes originating from the anomalous Hall effect (AHE) and the planar Hall effect (PHE), respectively. $B_{\text{DL}}$, $B_{\text{FL}}$, and $B_{\text{Oe}}$ represent the damping-like (DL) effective field, field-like (FL) effective field, and Oersted field, respectively. The term $I_c \alpha \nabla T$ in Eq. (3) takes into account the contributions of thermoelectric effects, *e.g.* anomalous Nernst effect and spin Seebeck effect, the thermoelectric coefficient ($\alpha$), and the thermal gradient along $z$ ($\nabla T$) as parameters. In Eq. (3), one sees that the DL torque component and the thermoelectric component are both proportional to $\cos\phi$, whereas the FL torque component is proportional to $(2\cos^3\phi - \cos\phi = \cos2\phi \cos\phi)$. Eq. (3) suggests, by measuring the $\phi$ dependence of $R^{2\omega}$, the FL torque contribution can be distinguished from those of the DL torque and thermoelectric effects. Defining the coefficients of $C$ and $D$, Eq. (3) can be rewritten as

$$R^{2\omega} = C\cos\phi + D(2\cos^3\phi - \cos\phi), \quad (4)$$

where $C = R_{\text{AHE}}\{B_{\text{DL}}/(B + B_{\text{ani}})\} + I_c\alpha\nabla T$, and $D = 2R_{\text{PHE}}\{(B_{\text{FL}} + B_{\text{Oe}})/B\}$. Further separation of DL torque component and thermoelectric component is possible from the $B$ dependence of $C$.

Figures 6(b) and 6(c) display the $\phi$ dependence of $R^{\omega}$ and $R^{2\omega}$, respectively, with $B = 0.1$ T for the Co (2 nm) / Cu$_{95}$Ir$_5$ (5 nm) bilayer. In Fig. 6(b), the fit to the $R^{\omega}$ values by $\sin2\phi$ function (red curve) led to $|R_{\text{PHE}}| = 2.97$ m$\Omega$. In Fig. 6(c), the orange open circles represent the measured data, and the red curve is the result of the fitting by $\cos\phi$. The green open circles denote the values upon subtracting the $\cos\phi$ component from the measured data, and the black curve is the best fit by the $\cos2\phi \cos\phi$ function. As seen in Fig. 6(c), the measured data of $R^{2\omega}$ are well fitted by the $\cos\phi$ function. In contrast to the original measured data, the green circles in Fig. 6(c) do not exhibit a clear angular dependence in amplitude following the $\cos2\phi \cos\phi$ function. Consequently, neither the size nor the sign of the FL torque can be securely determined from the fitted line (black curve). Figure 6(d) shows the Hall resistance ($R_{xy}$) as a function of $B$ applied along the direction perpendicular to the device plane, which corresponds to $\theta = 0$ or $\pi$, in which a dc current of 20 μA was applied. The $R_{xy}$ change is dominated by the AHE of the Co layer, and $R_{AHE}$ was obtained to be 0.30 Ω. The $B_{ani}$ was also estimated to be 0.75 T from Fig. 6(d). The $B_{Oe}$ was estimated to be -0.19 mT following the Ampère's law. The coefficient $C$ ($D$) as a function of the inverse of $B + B_{ani}$ (the inverse of $B$) is plotted in Fig. 6(e) (Fig. 6(f)). From the linear fits to the values of $C$ and $D$, $B_{DL}$ and $B_{FL}$ are estimated to be 0.15 mT and 0.19 mT, respectively. The spin Hall efficiencies can be estimated by the following equations [64]:

$$\xi_{DL}^{j} = \frac{2e}{\hbar} \frac{B_{DL} M_s t_{Co}}{j_{Cu-Ir}},$$
$$\xi_{FL}^{j} = \frac{2e}{\hbar} \frac{B_{FL} M_s t_{Co}}{j_{Cu-Ir}}, \quad (5)$$

where $e$ (< 0) is the elementary charge of an electron and $\hbar$ is the reduced Plank constant. $j_{Cu-Ir}$ is the current density flowing in the $Cu_{95}Ir_5$ layer. With the parameters of $M_s$ = 1200 kA/m and $j_{Cu-Ir}$ = 6.2 × 10$^6$ A/cm$^2$, $\xi_{DL}^{j}$ = 1.8 ± 0.1 % and $\xi_{FL}^{j}$ = 2.2 ± 0.2 % were obtained for the present Co (2 nm) / $Cu_{95}Ir_5$ (5 nm) bilayer. Considering the FM / NM bilayer structure (NM layer being on top of the FM layer) in this study, the positive $\xi_{DL}^{j}$ implies that the $\theta_{SH}$ of the $Cu_{95}Ir_5$ layer is positive (the same sign as that of Pt), which is consistent with Ref. [45]. A positive $\xi_{FL}^{j}$ in our convention then reflects that the FL torque is opposite to the Oersted field.

The accuracy in the estimation of $\xi_{FL}^{j}$ should be noted here. As explained above, the measured $R^{2\omega}$ did not clearly show the $\cos 2\phi \cos\phi$ dependence, which may be attributed to the small $R_{PHE}$. Therefore, we consider that the estimation of $\xi_{FL}^{j}$ had low accuracy. An important point is that, in contrast to $\xi_{FL}^{j}$, the estimation of $\xi_{DL}^{j}$ is not affected by the magnitude of $R_{PHE}$. Hereafter, we will mainly discuss the DL torque component, and evaluate $\theta_{SH}$ and $\lambda_{SD}$ of the $Cu_{95}Ir_5$ from $\xi_{DL}^{j}$.

The DL spin torque efficiency per unit applied electric field ($\xi_{DL}^{E}$) can be described

as [64]

$$\xi_{DL}^{E} \equiv \frac{2e}{\hbar} \frac{B_{DL} M_s t}{j_{Cu-Ir} E} = T_{int} \sigma_{SH} = \frac{\xi_{DL}^{j}}{\rho_{Cu-Ir}}, \quad (6)$$

where $E$ is the electric field in the NM layer. By considering the interface spin transparency ($T_{int}$), which is given by the ratio of {($B_{DL}M_s t$) / $j_{Cu-Ir}$} to the spin current originating from the bulk SHE ($j_{SHE}$) and is less than 1, $\xi_{DL}^{E}$ can be expressed as the product of $T_{int}$ and the spin Hall conductivity ($\sigma_{SH}$) defined by $(2e/\hbar)j_{SHE}/E$. In this study, we assume that the DL torque comes from the SHE in the NM layer through the well-ordered interface without spin memory loss [65] and $T_{int} < 1$ dominantly results from the spin back flow. Then, the following relation can be approximately expected [40,64,66] :

$$\xi_{DL}^{E}(t_{Cu-Ir}) = \frac{2e}{\hbar} \sigma_{SH}[1 - \text{sech}(t_{Cu-Ir}/\lambda_{SD})] \left(1 + \frac{\tanh(t_{Cu-Ir}/\lambda_{SD})}{2\lambda_{Cu-Ir}\rho_{Cu-Ir}\text{Re}[G_{MIX}]}\right)^{-1}. \quad (7)$$

From the fit to the $t_{Cu-Ir}$ dependence of $\xi_{DL}^{E}$ with Eq. (7), the $\sigma_{SH}$ and $\lambda_{SD}$ for the Cu$_{95}$Ir$_5$ are obtained, and the resultant $\theta_{SH}$ is also estimated.

Figure 7 (a) shows the $t_{Cu-Ir}$ dependence of $\rho_{Cu-Ir}$ for the Co (2 nm) / Cu$_{95}$Ir$_5$ ($t_{Cu-Ir}$) bilayer samples. $\rho_{Cu-Ir}$ shows the value around 90 μΩ cm regardless of $t_{Cu-Ir}$. This $\rho_{Cu-Ir}$ is consistent with the value obtained in Fig. 5(b). Figure 7(b) plots the $t_{Cu-Ir}$ dependence of $\xi_{DL}^{j}$. $\xi_{DL}^{j}$ saturates at around $t_{Cu-Ir}$ = 9 nm. The large error bars of $\xi_{DL}^{j}$ for $t_{Cu-Ir}$ = 15 nm and 20 nm are due to the large amount of shunting current into the Cu$_{95}$Ir$_5$ layer. Figure 7(c) is the $t_{Cu-Ir}$ dependence of $\xi_{DL}^{E}$. The red solid curve represents the fitting result to the experimental data using Eq. (7). With the parameters of $\rho_{Cu-Ir}$ = 92.42 μΩ cm and Re[$G_{MIX}$] = 1.0 × 10$^{15}$ Ω$^{-1}$m$^{-2}$, $\sigma_{SH}$ and $\lambda_{SD}$ of the Cu$_{95}$Ir$_5$ were estimated to be 3.2 × 10$^4$ [$\hbar/2e$]Ω$^{-1}$m$^{-1}$ and 2.9 ± 0.8 nm, respectively. Finally, the result of harmonic Hall voltage measurements suggested that $\theta_{SH}$ (= $\sigma_{SH} \rho_{Cu-Ir}$) is 3.0 ± 0.3 % for the present Cu$_{95}$Ir$_5$. Table 1 summarizes the values of $\theta_{SH}$ and $\lambda_{SD}$ evaluated using two different techniques. The difference between the two estimated values may be ascribed to the ambiguity of fitting parameters such as $P$ and $\lambda_{Co}$. In addition, other mechanisms including anomalous SMR which was reported in the Pt / Co bilayers [67] and anisotropic magnetoresistance in textured FM films [68], may also contribute to the observed SMR signal. Considering these points, we conclude that the present $Cu_{95}Ir_5$ possesses a relatively large $\theta_{SH}$ and a short $\lambda_{SD}$.

## 3.6. Discussion

First, let us discuss the relationship between $-J_{AF}$ through the $Cu_{95}Ir_5$ interlayer and the SHE and $\lambda_{SD}$ in the $Cu_{95}Ir_5$. As discussed in Sec. 3.2, the Co / $Cu_{95}Ir_5$ / Co exhibits a smaller $-J_{AF}$ than those for the Co / Cu / Co and Co / Ir / Co. On the other hand, the interlayer thickness showing $-J_{max}$ is $t = 0.75$ nm for the $Cu_{95}Ir_5$ interlayer sample, which is the same as that for the Cu interlayer sample. Hereafter we assume that 5 at.%-Ir doping does not remarkably change the shape of Fermi surface of pure Cu. Since the RKKY interaction and the interlayer exchange coupling in FM / NM / FM tri-layers may be regarded as the indirect exchange interaction between the local spins of the FMs, mediated via the conduction electron spins in the NM spacer, the transport mean free path ($l_{MFP}$) and the $\lambda_{SD}$ of the conduction electrons for the NM are also an important length scales. We experimentally found that the doping of 5 at.%-Ir significantly shortens the $\lambda_{SD}$ and increases the resistivity. The $\lambda_{SD} \leq 3$ nm for the present $Cu_{95}Ir_5$ is two order of magnitude shorter than that for pure Cu, *e.g.* $\lambda_{SD} \geq 300$ nm at room temperature for Cu [69]. Meanwhile, assuming $\rho_{Cu} \sim 10$ μΩ cm, Ir doping at the same level increases the resistivity only by nearly an order of magnitude. $\lambda_{SD}$ in a NM is related to $l_{MFP}$ and the spin mean free path, *i.e.* $\lambda_{SD} \propto \sqrt{l_{MFP} v_F \tau_{SF}}$, where $v_F$ and $\tau_{SF}$ are the Fermi velocity and spin flip time, respectively [70]. In addition, the relationship $l_{MFP} \propto 1/\rho$ holds according to Ref. [71]. Our experimental data indicate that the shortened $l_{MFP}$ alone cannot account for the drastic reduction of $\lambda_{SD}$ in Cu, upon doping with 5 at.% Ir. Instead, we consider introducing Ir dopant with strong spin-orbit coupling into a Cu lattice of weak spin-orbit coupling dramatically reduces the $\tau_{SF}$ of the latter by at least three order of magnitude. Since the optimum thickness for achieving the strongest coupling ($t \sim 0.75$ nm) represents a non-negligible fraction of the $\lambda_{SD}$ for $Cu_{95}Ir_5$, we consider the spin flip scattering within $Cu_{95}Ir_5$ spacer may be account for the reduction of -$J_{AF}$.

The $Cu_{95}Ir_5$ in this study showed $\theta_{SH} = 3.0 \pm 0.3$ % (4.3 %) and $\lambda_{SD} = 2.9 \pm 0.8$ nm (1.3 $\pm$ 0.1 nm) in the harmonic Hall voltage (SMR) measurement. Niimi *et al.* [45] reported the SHE in $Cu_{100-x}Ir_x$ with $0 \leq x \leq 12$ and obtained $\theta_{SH} = 2.1 \pm 0.6$ % and $\lambda_{SD} \sim 10$ nm at 10 K. Compared with those values, our $Cu_{95}Ir_5$ shows a larger $\theta_{SH}$ and a shorter $\lambda_{SD}$. Although at present the reason for the difference between the present study and the previous one is not definite, we believe that the dominant scattering mechanisms leading to the SHE in the two cases may be different. In the previous study [45], they claimed that the skew scattering plays a major role for the SHE. However, we consider that our $Cu_{95}Ir_5$ is out of the composition range that the skew scattering is dominant because of the low conductivity ($\sigma \sim 1.1 \times 10^4$ $\Omega^{-1}$ cm$^{-1}$). Instead, the side jump process or the intrinsic process may be dominant for our $Cu_{95}Ir_5$ because $\sigma \sim 1.1 \times 10^4$ $\Omega^{-1}$ cm$^{-1}$ is in the moderately dirty region of $10^3 \leq \sigma \leq 10^5$ $\Omega^{-1}$ cm$^{-1}$ [72]. In the intrinsic region, it is likely that there exists a trade-off between $\lambda_{SD}$ and $\theta_{SH}$ for $Cu_{100-x}Ir_x$, as in the case of Pt, such that the product of $\lambda_{SD}$ and $\theta_{SH}$ is nearly a constant [65,73]. This possible scenario may reconcile our results with previous observations.

The striking feature of this study is that the Ir-doped Cu can exhibit large SHE comparable to those for the typical spin Hall materials such as Pt. Since the Pt has already been utilized as a spin current source for SOT switching [10,12], we anticipate that the $Cu_{95}Ir_5$ interlayer is one of the promising nonmagnetic materials allowing us to demonstrate the SOT switching in the artificial metallic superlattices with the AF-coupling. In addition, provided that the SHE of $Cu_{100-x}Ir_x$ can be further enhanced, *e.g.* by exploiting the interfacial spin current at FM / $Cu_{100-x}Ir_x$ interface [74] and/or by combining with the large SHE of an appropriate FM [75,76], the alloy will be an interesting candidate for achieving efficient SOT switching in FM / spacer / perpendicular FM free layer structure, without the need an external magnetic field [77], which is a prerequisite for next-generation SOT-Magnetic Random Access Memory (MRAM).

## 4. Summary

We have investigated the interlayer exchange coupling for the Co (2 nm) / X ($t$) / Co (2 nm) systems, where X = Cu, $Cu_{95}Ir_5$, or Ir. The AF-coupling was observed in the magnetization curve even for the $Cu_{95}Ir_5$ interlayer samples. From the $t$ dependence of the $J_{AF}$, the value of -$J_{max}$ were obtained to be 0.39 mJ/m$^2$ at $t$ = 0.75 nm for the $Cu_{95}Ir_5$ interlayer sample. We have also evaluated the SHE for the $Cu_{95}Ir_5$ using two different evaluation methods: the SMR and the harmonic Hall voltage measurements. Finally, a large $\theta_{SH}$ = 3.0 ± 0.3 % (4.3 %) and a short $\lambda_{SD}$ = 2.9 ± 0.8 nm (1.3 ± 0.1 nm) were obtained for $Cu_{95}Ir_5$ from the harmonic Hall voltage (SMR) measurement. This study has clarified that $Cu_{95}Ir_5$ is a promising nonmagnetic spacer layer simultaneously showing interlayer exchange coupling and relatively large SHE. We expect the $Cu_{95}Ir_5$ interlayer will pave a new way to the AF spintronics based on the metallic superlattice.

**Acknowledgements**

The authors express our gratitude to I. Narita for his technical support for the composition analysis. The structural characterization and the device fabrication were partly carried out at the Cooperative Research and Development Center for Advanced Materials, IMR, Tohoku University. This work was supported by the Grant-in-Aid for Scientific Research (S) (JP18H05246) from JSPS KAKENHI, Japan.

# References

[1] T. Jungwirth, X. Marti, P. Wadley, and J. Wunderlich, Nat. Nanotechnol. **11**, 231-241 (2016).

[2] V. Baltz, A. Manchon, M. Tsoi, T. Moriyama, T. Ono, and Y. Tserkovnyak, Rev. Mod. Phys. **90**, 015005 (2018).

[3] Z. Wei, A. Sharma, A. S. Nunez, P. M. Haney, R. A. Duine, J. Bass, A. H. MacDonald, and M. Tsoi, Phys. Rev. Lett. **98**, 116603 (2007).

[4] L. Néel, http://www.nobelprize.org/nobel_prizes/physics/laureates/1970/neel-lecture.pdf.

[5] D. Houssameddine, J. F. Sierra, D. Gusakova, B. Delaet, U. Ebels, L. D. Buda-Prejbeanu, M. C. Cyrille, B. Dieny, B. Ocker, J. Langer, and W. Maas, Appl. Phys. Lett. **96**, 072511 (2010).

[6] T. Seki, H. Tomita, M. Shiraishi, T. Shinjo, and Y. Suzuki, Appl. Phys. Exp. **97**, 033001 (2010).

[7] T. Seki, H. Tomita, T. Shinjo, and Y. Suzuki, Appl. Phys. Lett. **97**, 162508 (2010).

[8] K. Tanaka, T. Moriyama, M. Nagata, T. Seki, K. Takanashi, S. Takahashi, and T. Ono, Appl. Phys. Exp. **7**, 063010 (2014).

[9] P. Wadley, B. Howells, J. Železný, C. Andrews, V. Hills, R. P. Campion, V. Novák, K. Olejník, F. Maccherozzi, S. S. Dhesi, S. Y. Martin, T. Wagner, J. Wunderlich, F. Freimuth, Y. Mokrousov, J. Kuneš, J. S. Chauhan, M. J. Grzybowski, A. W. Rushforth, K. Edmond, B. L. Gallagher, and T. Jungwirth, Science **351**, 6273, 587-590 (2016).

[10] T. Moriyama, K. Oda, T. Ohkochi, M. Kimata, and T. Ono, Sci. Rep. **8**, 14167 (2018).

[11] S. Y. Bodnar, L. Šmejkal, I. Turek, T. Jungwirth, O. Gomonay, J. Sinova, A. A. Sapozhnik, H. J. Elmers, M. Kläui, and M. Jourdan, Nat. Commun. **9**, 348 (2018).

[12] T. Moriyama, W. Zhou, T. Seki, K. Takanashi, and T. Ono, Phys. Rev. Lett. **121**, 167202 (2018).

[13] X. Z. Chen, R. Zarzuela, J. Zhang, C. Song, X. F. Zhou, G. Y. Shi, F. Li, H. A. Zhou, W. J. Jiang, F. Pan, and Y. Tserkovnyak, Phys. Rev. Lett. **120**, 207204 (2018).

[14] W. Zhou, T. Seki, T. Kubota, G. E. W. Bauer, and K. Takanashi, Phys. Rev. Mater. **2**, 094404 (2018).

[15] M. Meinert, D. Graulich, and T. Matalla-Wagner, Phys. Rev. Appl. **9**, 064040 (2018).

[16] M. J. Grzybowski, P. Wadley, K. W. Edmonds, R. Beardsley, V. Hills, R. P. Campion, B. L. Gallagher, J. S. Chauhan, V. Novak, T. Jungwirth, F. Maccherozzi, and S. S. Dhesi, Phys. Rev. Lett. **118**, 057701 (2017).

[17] A. A. Sapozhnik, M. Filianina, S. Y. Bodnar, A. Lamirand, M. A. Mawass, Y. Skourski, H. J. Elmers, H. Zabel, M. Kläui, and M. Jourdan, Phys. Rev. B **97**, 134429 (2018).

[18] H. Nakayama, M. Althammer, Y.-T. Chen, K. Uchida, Y. Kajiwara, D. Kikuchi, T. Ohtani, S. Geprägs, M. Opel, S. Takahashi, R. Gross, G. E. W. Bauer, S. T. B. Goennenwein, and E. Saitoh, Phys. Rev. Lett. **110**, 206601 (2013).

[19] C. O. Avci, K. Garello, M. Gabureac, A. Ghosh, A. Fuhrer, S. F. Alvarado, and P. Gambardella, Phys. Rev. B **90**, 224427 (2014).

[20] Y.-C. Lau and M. Hayashi, Jpn. J. Appl. Phys. **56**, 0802B5 (2017).

[21] J. Kim, P. Sheng, S. Takahashi, S. Mitani, and M. Hayashi, Phys. Rev. Lett. **116**, 097201 (2016).

[22] C. Hahn, G. De Loubens, O. Klein, M. Viret, V. V. Naletov, and J. Ben Youssef, Phys. Rev. B **87**, 174417 (2013).

[23] C. F. Pai, L. Liu, Y. Li, H. W. Tseng, D. C. Ralph, and R. A. Buhrman, Appl. Phys. Lett. **101**, 122404 (2012).

[24] L. Liu, C. F. Pai, Y. Li, H. W. Tseng, D. C. Ralph, and R. A. Buhrman, Science **336**, 6081, 555-558 (2012).

[25] P. Grünberg, R. Schreiber, Y. Pang, U. Walz, M. B. Brodsky, and H. Sowers, Phys. Rev. Lett. **57**, 2442 (1986).

[26] C. Carbone and S. F. Alvarado, Phys. Rev. B **36**, 2433(R) (1987).

[27] D. H. Mosca, F. Petroff, A. Fert, P. A. Schroeder, W. P. Pratt, and R. Laloee, J. Magn. Magn. Mater. **94**, 1-2 (1991).

[28] S. S. P. Parkin, R. F. Marks, R. F. C. Farrow, G. R. Harp, Q. H. Lam, and R. J. Savoy, Phys. Rev. B **46**, 9242(R) (1992).

[29] M. A. Howson, B. J. Hickey, J. Xu, D. Greig, and Nathan Wiser, Phys. Rev. B **48**, 1322(R) (1993).

[30] J. P. Renard, P. Beauvillain, C. Dupas, K. Le Dang, P. Veillet, E. Vélu, C. Marlière, and D. Renard, J. Magn. Magn. Mater. **115**, 2-3, L147-L151 (1992).

[31] C. Dupas, E. Kolb, K. Le Dang, J. P. Renard, P. Veillet, E. Vélu, and D. Renard, J. Magn. Magn. Mater. **128**, 361-364 (1993).

[32] H. Yanagihara, E. Kita, and M. B. Salamon, Phys. Rev. B **60**, 12957 (1999).

[33] Y. Luo, M. Moske, and K. Samwer, Europhys. Lett. **42**, 565 (1998).

[34] A. Dinia, M. Stoeffel, K. Rahmouni, D. Stoeffler, and H. A. M. Van Den Berg, Europhys. Lett. **42**, 331 (1998).

[35] C. F. Majkrzak, J. W. Cable, J. Kwo, M. Hong, D. B. McWhan, Y. Yafet, J. V. Waszczak, and C. Vettier, Phys. Rev. Lett. **57**, 923 (1986).

[36] M. B. Salamon, S. Sinha, J. J. Rhyne, J. E. Cunningham, R. W. Erwin, J. Borchers, and C. P. Flynn, Phys. Rev. Lett. **56**, 259 (1986).

[37] M. T. Johnson, R. Coehoorn, J. J. de Vries, N. W. E. McGee, J. aan de Stegge, and P. J. H. Bloemen, Phys. Rev. Lett. **69**, 969 (1992).

[38] J. Kohlhepp, S. Cordes, H. J. Elmers, and U. Gradmann, J. Magn. Magn. Mater. **111**, 3, L231-L234 (1992).

[39] A. Schreyer, K. Bröhl, J. F. Ankner, C. F. Majkrzak, Th. Zeidler, P. Bödeker, N. Metoki, and H. Zabel, Phys. Rev. B **47**, 15334(R) (1993).

[40] P. M. Haney, H. W. Lee, K. J. Lee, A. Manchon, and M. D. Stiles, Phys. Rev. B **88**, 214417 (2013).

[41] L. Zhu, D. C. Ralph, and R. A. Buhrman, Phys. Rev. Lett. **122**, 077201 (2019).

[42] L. Wang, R. J. H. Wesselink, Y. Liu, Z. Yuan, K. Xia, and P. J. Kelly, Phys. Rev. Lett. **116**, 196602 (2016).

[43] Y.-C. Lau, Z. Chi, T. Taniguchi, M. Kawaguchi, G. Shibata, N. Kawamura, M. Suzuki, S. Fukami, A. Fujimori, H. Ohno, and M. Hayashi, Phys. Rev. Mater. **3**, 104419 (2019).

[44] Y. Ishikuro, M. Kawaguchi, N. Kato, Y. C. Lau, and M. Hayashi, Phys. Rev. B **99**, 134421 (2019).

[45] Y. Niimi, M. Morota, D. H. Wei, C. Deranlot, M. Basletic, A. Hamzic, A. Fert, and Y. Otani, Phys. Rev. Lett. **106**, 126601 (2011).

[46] M. Yamanouchi, L. Chen, J. Kim, M. Hayashi, H. Sato, S. Fukami, S. Ikeda, F. Matsukura, and H. Ohno,

Appl. Phys. Lett. **102**, 212408 (2013).

[47] J. Cramer, T. Seifert, A. Kronenberg, F. Fuhrmann, G. Jakob, M. Jourdan, T. Kampfrath, and M. Kläui, Nano Lett. **18**, 2, 1064-1069 (2018).

[48] S. Takizawa, M. Kimata, Y. Omori, Y. Niimi, and Y. Otani, Appl. Phys. Express **9**, 063009 (2016).

[49] S. Demokritov, E. Tsymbal, P. Grünberg, W. Zinn, and Ivan K. Schuller, Phys. Rev. B **49**, 720(R) (1994).

[50] J. C. Slonczewski, Phys. Rev. Lett. **67**, 3172 (1991).

[51] J. C. Slonczewski, J. Magn. Magn. Mater. **150**, 13-24 (1995).

[52] S. Bosu, Y. Sakuraba, K. Saito, K. Izumi, T. Koganezawa, and K. Takanashi, J. Magn. Magn. Mater. **369**, 211-218 (2014).

[53] P. Grunberg et al., *Handbook of Magnetic Materials, Volume 13* (North Holland, 2001).

[54] P. Bruno and C. Chappert, Phys. Rev. Lett. **67**, 1602 (1991).

[55] A. Brataas, Y. V. Nazarov, and G. E. W. Bauer, Phys. Rev. Lett. **84**, 2481 (2000).

[56] M. Weiler, M. Althammer, M. Schreier, J. Lotze, M. Pernpeintner, S. Meyer, H. Huebl, R. Gross, A. Kamra, J. Xiao, Y. T. Chen, H. Jiao, G. E. W. Bauer, and S. T. B. Goennenwein, Phys. Rev. Lett. **111**, 176601 (2013).

[57] J. Kim, J. Sinha, S. Mitani, M. Hayashi, S. Takahashi, S. Maekawa, M. Yamanouchi, and H. Ohno, Phys. Rev. B **89**, 174424 (2014).

[58] M. D. Stiles and A. Zangwill, Phys. Rev. B **66**, 0144071 (2002).

[59] L. Zhu, D. C. Ralph, and R. A. Buhrman, Phys. Rev. Lett. **123**, 057203 (2019).

[60] P. M. Tedrow and R. Meservey, Phys. Rev. B **7**, 318-326 (1973).

[61] M. B. Stearns, J. Magn. Magn. Mater. **5**, 167-171 (1977).

[62] G. Zahnd, L. Vila, V. T. Pham, M. Cosset-Cheneau, W. Lim, A. Brenac, P. Laczkowski, A. Marty, and J. P. Attané, Phys. Rev. B **98**, 174414 (2018).

[63] J. S. Moodera, L. R. Kinder, T. M. Wong, and R. Meservey, Phys. Rev. Lett. **74**, 3273 (1995).

[64] M.-H. Nguyen, D. C. Ralph, and R. A. Buhrman, Phys. Rev. Lett. **116**, 126601 (2016).

[65] J. C. Rojas-Sánchez, N. Reyren, P. Laczkowski, W. Savero, J. P. Attané, C. Deranlot, M. Jamet, J. M. George, L. Vila, and H. Jaffrès, Phys. Rev. Lett. **112**, 106602 (2014).

[66] Y. T. Chen, S. Takahashi, H. Nakayama, M. Althammer, S. T. B. Goennenwein, E. Saitoh, and G. E. W. Bauer, Phys. Rev. B **87**, 144411 (2013).

[67] M. Kawaguchi, D. Towa, Y.-C. Lau, S. Takahashi, and M. Hayashi, Appl. Phys. Lett. **112**, 202405 (2018).

[68] A. Philippi-Kobs, A. Farhadi, L. Matheis, D. Lott, A. Chuvilin, and H. P. Oepen, Phys. Rev. Lett. **123**, 137201 (2019).

[69] F. J. Jedema, A. T. Filip and B. J. van Wees, Nature **410**, 345-348 (2001).

[70] T. Valet and A. Fert, Phys. Rev. B **48**, 7099 (1993).

[71] O. Gunnarsson, M. Calandra, and J. E. Han, Rev. Mod. Phys. **75**, 1085 (2003).

[72] S. Onoda, N. Sugimoto, and N. Nagaosa, Phys. Rev. B **77**, 165103 (2008).

[73] E. Sagasta, Y. Omori, M. Isasa, M. Gradhand, L. E. Hueso, Y. Niimi, Y. Otani, and F. Casanova, Phys. Rev. B **94**, 060412(R) (2016).

[74] Seung-heon C. Baek, V. P. Amin, Y.-W. Oh, G. Go, S.-J. Lee, G.-H. Lee, K. J. Kim, M. D. Stiles, B.-G.

[75]   Park, and K.-J. Lee, Nat. Mater. **17**, 509-513 (2018).

[75]   H. Iihama, S., Taniguchi, T., Yakushiji, K., Fukushima, A., Shiota, Y., Tsunegi, S., Hiramatsu, R., Yuasa, S., Suzuki, Y. & Kubota, Nat. Electron. **1**, 120-123 (2018).

[76]   T. Seki, S. Iihama, T. Taniguchi, and K. Takanashi, Phys. Rev. B **100**, 144427 (2019).

[77]   Y.-C. Lau, D. Betto, K. Rode, J. M. D. Coey, and P. Stamenov, Nat. Nanotechnol. **11**, 758-762 (2016).

**Table 1: The values of $\theta_{SH}$ and $\lambda_{SD}$ of the Cu$_{95}$Ir$_5$ evaluated by the SMR measurement and the harmonic Hall voltage measurement.**

| Evaluation method | $\theta_{SH}$ (%) | $\lambda_{SD}$ (nm) |
|:---:|:---:|:---:|
| SMR | 4.3 | 1.3 ± 0.1 |
| Harmonic Hall voltage | 3.0 ± 0.3 | 2.9 ± 0.8 |

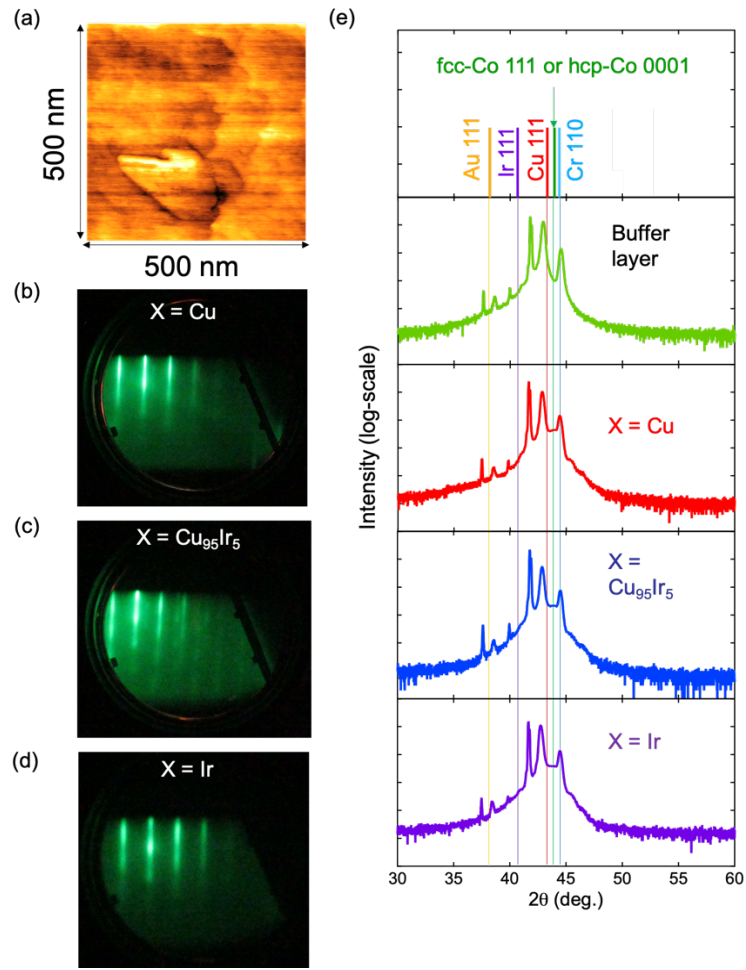

Fig. 1: (a) AFM image for the buffer layer consisting of Cr (10 nm) / Au (5 nm) / Cu (35 nm)). (b)-(d) The RHEED images for (b) Cu, (c) $Cu_{95}Ir_5$, and (d) Ir interlayer samples, respectively, taken after the deposition of top 2 nm-thick Co layer. (e) Out-of-plane XRD profiles for the powder diffraction data of the bulk Au 111, Ir 111, Cu 111, fcc Co 111 or hcp Co 0001, and Cr 110, the buffer layer only sample (light green curve), the Cu interlayer sample (red curve), the $Cu_{95}Ir_5$ interlayer sample (blue curve), and the Ir interlayer sample (purple curve).

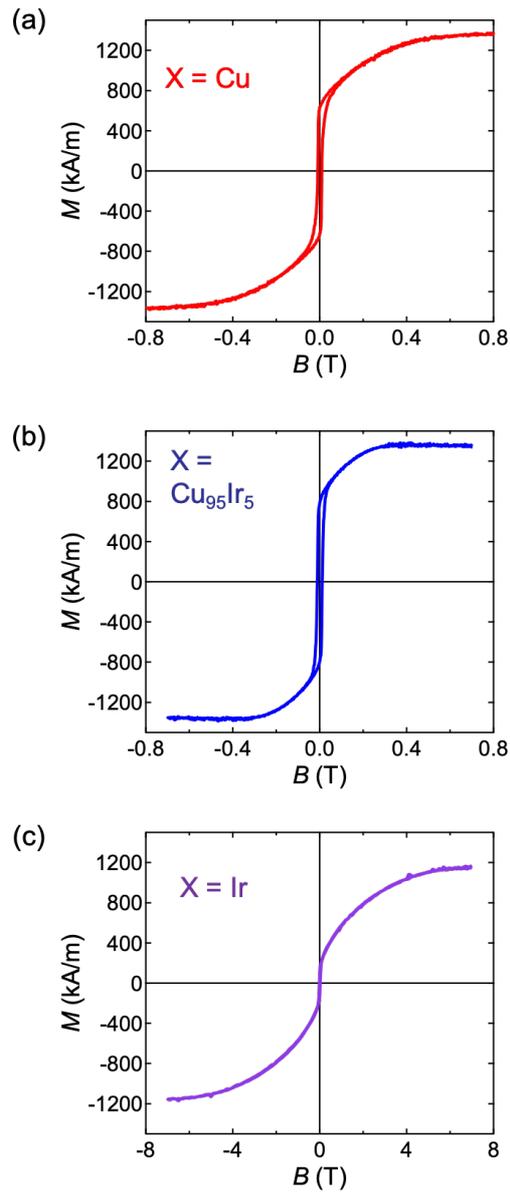

**Fig. 2:** Magnetization curves for (a) the Cu interlayer sample, (b) the $Cu_{95}Ir_5$ interlayer sample, and (c) the Ir interlayer samples. The $B$ was applied along the film plane for all the samples.

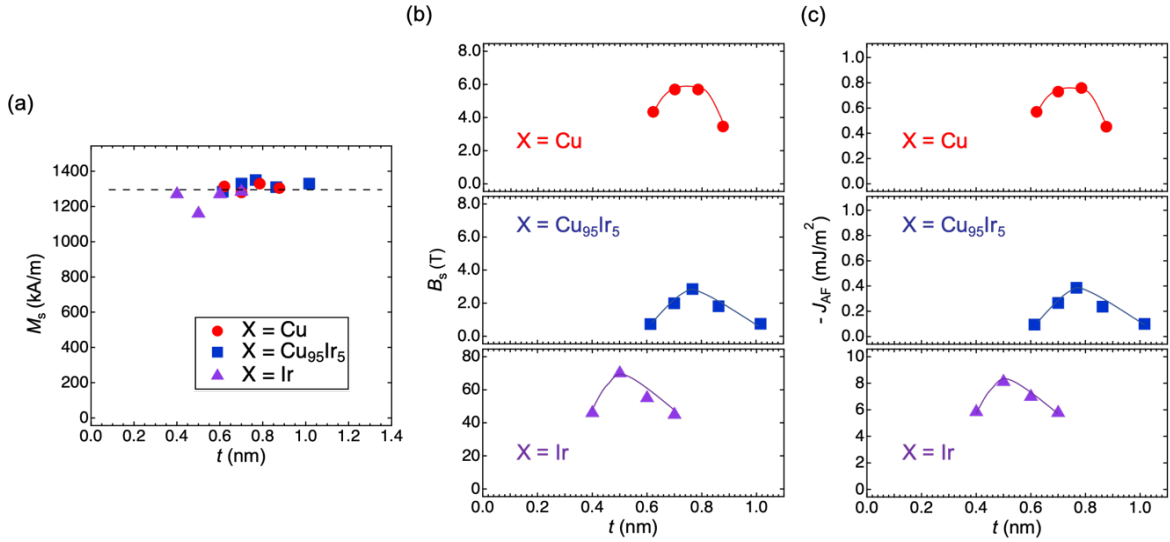

Fig. 3: $t$ dependence of (a) $M_s$, (b) $B_s$, and (c) $-J_{AF}$ obtained from the magnetization curves. The dotted line in (a) denote the magnetization value of 1300 kA/m. The red circles, blue squares, and purple triangles represent the results of the Cu interlayer sample, the $Cu_{95}Ir_5$ interlayer sample, and the Ir interlayer sample, respectively. The solid curves in (b) and (c) are guides for the eye.

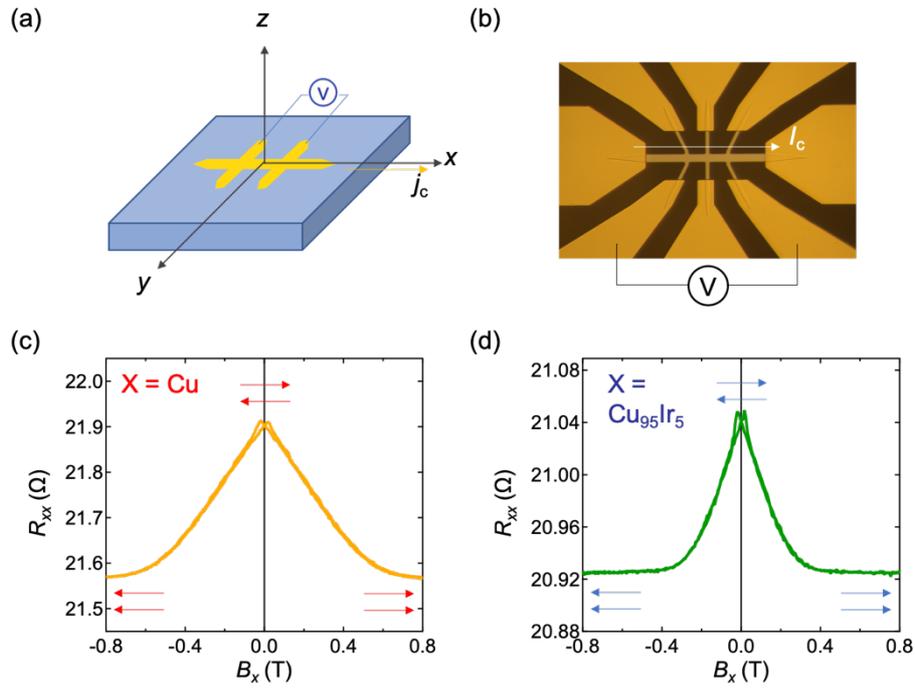

**Fig. 4: (a) Schematic illustration of the setup and the coordinate for the MR measurements. (b) Optical microscope image of Hall bar. (c) $R_{xx}$ for the Cu interlayer sample and (d) the $Cu_{95}Ir_5$ interlayer sample as a function of external magnetic field applied along the in-plane *x* direction ($B_x$).**

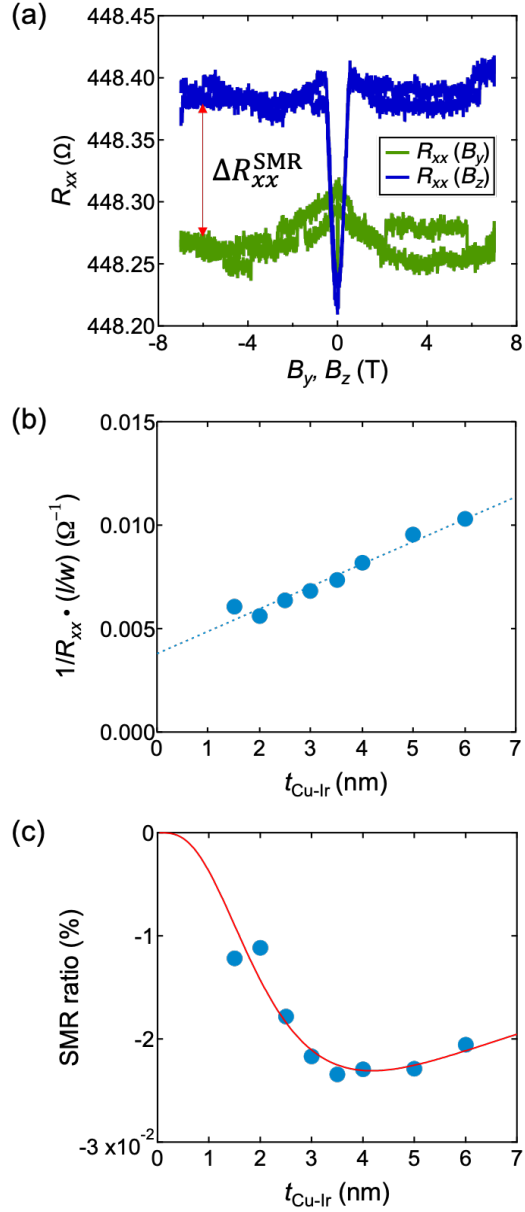

**Fig. 5 (a)** $R_{xx}$ - $B$ **curves for Co (2 nm) / Cu$_{95}$Ir$_5$ (2 nm) bilayer with the external magnetic field applied along the $y$ direction ($B_y$, green curve) and along the $z$ direction ($B_z$, blue curve). (b) $t_{Cu\text{-}Ir}$ dependence of inverse of sheet resistance (1 / $R_{xx}$ ($l/w$)). The blue dotted line represents the linear fit to the experimental data. (c) $t_{Cu\text{-}Ir}$ dependence of the SMR ratio. The red curve denotes the fit to the experimental data by Eq. (1).**

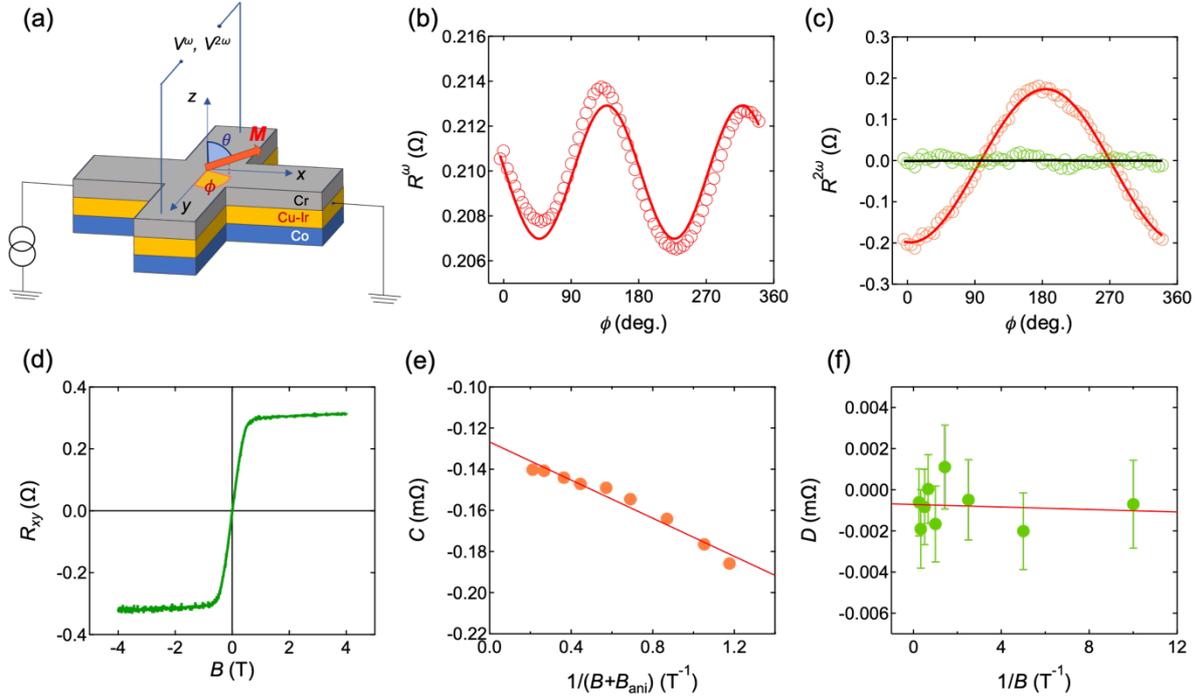

Fig. 6: (a) Illustration of the setup of harmonic Hall voltage measurement. (b) $\phi$ dependence of $R^{\omega}$ and (c) $R^{2\omega}$ with $B = 0.1$ T for the Al$_2$O$_3$ (0001) subs. / Co (2 nm) / Cu$_{95}$Ir$_5$ (5 nm) / Cr (5 nm). In (b), the red open circles denote the measured data, and the red solid curve represents the result of $\sin\phi$ fitting. In (c), the orange open circles represent the measured data, and the red solid curve is the result of fitting by $\cos\phi$. The green open circles represent the values, where the $\cos\phi$ component was subtracted from the measurement data. The black solid curve is the result of fitting by the $\cos 2\phi \cos\phi$ function. (d) Anomalous Hall loop with $B$ applied perpendicular to the device plane. (e) Coefficients $C$ and (f) $D$ of $R^{2\omega}$ as a function of the inverse of $B + B_{ani}$ and the inverse of $B$, respectively. The red solid lines in (e) and (f) are the results of linear fits.

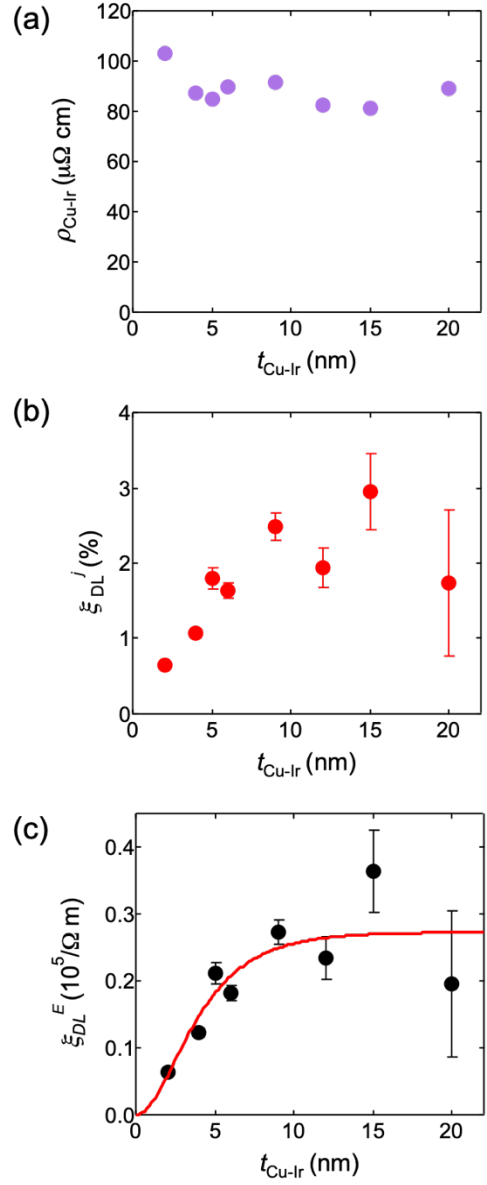

**Fig. 7:** $t_{\text{Cu-Ir}}$ dependence of (a) $\rho_{\text{Cu-Ir}}$, (b) $\xi_{DL}^{j}$, and (c) $\xi_{DL}^{E}$ obtained by the harmonic Hall voltage measurement. In (c), red solid curve denotes the result of fitting line with Eq. (7).